# Bias voltage controlled magnetization switch in ferromagnetic semiconductor resonant tunneling diodes

Swaroop Ganguly,[1,*] L.F.Register,[1] S.Banerjee,[1] and A.H.MacDonald[2]

[1]Microelectronics Research Center, University of Texas at Austin, TX 78758, USA.

[2]Department of Physics, University of Texas at Austin, TX 78712, USA.

We predict that the Curie temperature of a ferromagnetic resonant tunneling diode will decrease abruptly, by approximately a factor of two, when the downstream chemical potential falls below the quantum well resonance energy. This property follows from elementary quantum transport theory notions combined with a mean-field description of diluted magnetic semiconductor ferromagnetism. We illustrate this effect by solving coupled non-equilibrium Green's function, magnetic mean-field, and electrostatic Poisson equations self-consistently to predict the bias voltage and temperature dependence of the magnetization of a model system.

PACS numbers: 85.75.Mm, 72.25.Dc, 73.23.Ad

## I. INTRODUCTION

Diluted magnetic semiconductors (DMS's) that exhibit carrier-mediated ferromagnetism[1,2] have attracted interest because of new basic science questions that their properties raise and because of their potential for applications. In this article we predict that the magnetic properties of a resonant tunneling diode (RTD) with a DMS ferromagnet[3] quantum well, schematically illustrated in figure 1(a), will change abruptly when the down stream chemical potential crosses the quantum well resonance energy. The qualitative physics which leads to this effect is summarized by figure 1(b). In a mean-field description, whose approximate validity in strongly metallic DMS ferromagnets is now well established,[4,5] the ferromagnetic transition temperature of the quantum well system is determined by a competition between the entropic cost of moment ordering and the exchange energy gained by aligning band spins opposite to local moment spins. As we explain below, the Curie temperature is proportional to the spin-susceptibility of the band electrons, and therefore proportional to the energy gained by the paramagnetic band spin system when placed in an effective Zeeman field. In figure 1(b), the Zeeman field is characterized by the splitting $\Delta$ between majority and minority spin bands to which it gives rise. Standard transport theory assumptions imply that state occupation numbers in the presence of a bias voltage are obtained by averaging the Fermi factors of the up-stream and down-stream reservoirs of the RTD, with equal weight if the upstream and downstream barriers are identical. As illustrated by the arrows in figure 1(b), the gain in energy due to the Zeeman field can be thought of as following from the replacement of minority spin states in the energy interval (0, $\Delta/2$) by majority spin states in the interval ($-\Delta/2$,0), with all energies measured from the quantum well resonance (sub-band) energy. Since these states have occupation number 1 when the resonance energy is below the chemical potential of both reservoirs and 1/2 when the resonance energy is above the downstream chemical potential, the exchange energy gain is halved at this crossing. In an ideal disorder-free system with a narrow resonance, the Curie temperature will therefore drop by a factor of two.

This article is an examination of the idea explained above. In Section II we derive an analytic expression for the Curie temperature implied by the argument of figure 1. The simplifying assumptions used in this derivation are partially tested in Section III by performing a transport calculation using the non-equilibrium Greens function (NEGF) method,[6,7] treating electrostatics self-consistently by using the Poisson equation to calculate the electrostatic potential profile, and the quantum well magnetization self-consistently using mean field theory.[4,5] For definiteness we use material parameters that approximate a GaAs/AlGaAs/(Ga,Mn)As/AlGaAs/GaAs ferromagnetic RTD. We have simplified the problem by assuming a single isotropic and parabolic valence band with the heavy hole mass. In Section IV we present our conclusions and discuss some of the difficulties that stand in the way of the experimental realization of this effect.

## II. ANALYTICAL THEORY

When disorder within the quantum wells is neglected, RTD transport physics becomes one dimensional. For each transverse channel labeled by a two-dimensional wavevector $\vec{\kappa}$, the quantum well can be regarded as a single lattice site with energy $\mathrm{E} = E + \varepsilon_{\vec{\kappa}}$ (where $\varepsilon_{\vec{\kappa}} = \dfrac{\hbar^2 \vec{\kappa}^2}{2m^*}$ and $E$ is the quantum well resonance energy), that is coupled to reservoirs on the left and the right. In a mean-field description the quantum well resonance energy of up and down spin states are shifted when the Mn moments in the quantum well are polarized, *i.e.* when the DMS system is in its ferromagnetic state: $E_{\downarrow,\uparrow} = E \pm \Delta/2$ for minority and majority spin orbitals where the band spin-splitting[8]

$$\Delta \approx \int h_{pd}(z) \left|\varphi(z)\right|^2 dz = \int J_{pd} N_{Mn}(z) \langle M \rangle (z) \left|\varphi(z)\right|^2 dz . \qquad (1)$$

In equation (1), $\varphi(z)$ denotes the lowest sub-band quantum well wavefunction. For the approximate analytic calculations described in this section we make a narrow well, high barrier approximation by assuming that $\varphi(z)$ is independent of both bias voltages and $\Delta$. Here, $h_{pd}$ is the local kinetic exchange splitting, $J_{pd}$ is the strength of the exchange interaction between p-like valence band electrons and the d-shell local moments discussed later, $N_{Mn}$ is the density profile of the Mn local moments in the quantum well, and $\langle M \rangle$ is the mean spin-quantum number of the polarized $S = 5/2$ moments. The following argument generalizes the mean-field theory for $M$ from the case of electronic equilibrium to the non-equilibrium case. As we explain below the electronic property that enters the mean-field theory for the critical temperature is the band (Pauli) spin-susceptibility. We now show that the spin-susceptibility is suddenly decreased when the downstream chemical potential drops below the quantum well resonance energy.

The description of a non-equilibrium state in which two reservoirs with different chemical potentials are coupled to a single atomic site, and current flows from the high chemical potential reservoir to the low-chemical potential reservoir through that site, is the simplest toy model of quantum transport theory. Our interest here is in determining the influence of a bias voltage (i.e. a difference between the chemical potentials of the reservoirs) on the magnetic state of a DMS quantum well. The key result from quantum transport theory that we will need is the following simple expression[7] for the steady-state occupation of a quantum well state:

$$N = \frac{\gamma_L}{\gamma_L + \gamma_R} f_L(E) + \frac{\gamma_R}{\gamma_L + \gamma_R} f_R(E) \tag{2}$$

where, $f_{L,R} = \dfrac{1}{\exp\left(\left(E - \mu_{L,R}\right)/k_B T\right) + 1}$ are the Fermi functions for the left and right leads, and $\gamma_L/\hbar$, $\gamma_R/\hbar$ are the rates at which an electron inside the device will escape into the left and right leads, respectively. We use this expression to find the electronic charge and spin densities in the RTD quantum well. The couplings $\gamma_{L,R} \propto T_{L,R}$ where $T_{L,R} << 1$ are the tunneling probability through the left and right barriers.[9] For each spin subsystem, we integrate over transverse kinetic energies and assume a single longitudinal resonance level in the energy range of interest. We also assume degenerate statistics and parabolic dispersion. Under these conditions, the 2D carrier density in the well is given by the following expressions when $M = 0$:

$$N = 2 \cdot \nu \cdot \left[ \frac{\gamma_L}{\gamma_L + \gamma_R} \cdot \left(\mu_L - E\right) + \frac{\gamma_R}{\gamma_L + \gamma_R} \cdot \left(\mu_R - E\right) \right] \text{ for } \mu_L > \mu_R > E \tag{3}$$

$$N = 2 \cdot \nu \cdot \left[ \frac{\gamma_L}{\gamma_L + \gamma_R} \cdot \left(\mu_L - E\right) \right] \text{ for } \mu_L > E > \mu_R \tag{4}$$

where $\nu = \frac{m^*}{2\pi\hbar^2}$ is the 2D density of states and the factor of 2 accounts for spin degeneracy.

For tall nearly symmetric barriers, that is $\gamma_L \approx 1/2 \approx \gamma_R$, and a small applied bias $V$, with respect to the band-edge in the emitter side, $\mu_L = \mu$, $\mu_R = \mu - V$ by definition, and $E \approx \varepsilon - V/2$ where $\varepsilon$ is the resonance energy level at equilibrium, and all energies are measured in electron volts. Now, for $\mu_R > E \Rightarrow \mu - V > \varepsilon - V/2 \Rightarrow V < 2(\mu - \varepsilon)$, we have from equation (3):

$$N = 2 \cdot \nu \cdot \left[ (\mu - \varepsilon) + \frac{\gamma_L - \gamma_R}{\gamma_L + \gamma_R} \cdot \frac{V}{2} \right] \approx 2 \cdot \nu (\mu - \varepsilon) \tag{5}$$

Therefore, to lowest order in $\varepsilon$, the 2D spin susceptibility has the same value as in the equilibrium case:

$$\chi \cong \lim_{(\varepsilon_\downarrow - \varepsilon_\uparrow) \to 0} \frac{N_\uparrow - N_\downarrow}{\varepsilon_\downarrow - \varepsilon_\uparrow} \approx \nu \Rightarrow \frac{\chi}{\nu} \approx 1 \tag{6}$$

On the other hand, for $E > \mu_R \Rightarrow V > 2(\mu - \varepsilon)$, we get from equation (4):

$$N = 2 \cdot \nu \cdot \left[ \frac{\gamma_L}{\gamma_L + \gamma_R} \cdot (\mu - \varepsilon + V/2) \right] \approx \nu (\mu - \varepsilon + V/2) \tag{7}$$

which yields, for the 2D spin susceptibility:

$$\frac{\chi}{\nu} \approx \frac{\gamma_L}{\gamma_L + \gamma_R} \approx \frac{1}{2} \tag{8}$$

We note, from equations (5) and (7) that the total carrier concentration $N$ is continuous at the applied voltage $V = V_S \cong 2(\mu - \varepsilon)$, even though the spin susceptibility exhibits a sharp drop at this point as implied by equations (6) and (8). Finally, when $E < 0 \Rightarrow \varepsilon - V/2 < 0 \Rightarrow V > 2\varepsilon$, obviously $N \to 0 \Rightarrow \chi \to 0$. Therefore when the bias potential goes to $V_C \cong 2\varepsilon$, we expect the total concentration to decrease sharply and the spin susceptibility to vanish. Thus, the spin susceptibility is predicted to decrease with increasing voltage to zero from its equilibrium value in two steps, at $V_S$ and $V_C$. This qualitative argument is supported by numerical model simulations illustrated in figure 2 and described below.

The mean-field critical temperature in equilibrium DMS systems can be found by equating the Zeeman mean-field experienced by band electrons $h$ due to their interaction with induced local moment polarization, with the mean-field necessary to create a small spin density $\langle s \rangle = (n_\uparrow - n_\downarrow)/2$. Inserting the text book expression for local moment spin-susceptibility, one obtains:[5]

$$h = \frac{J_{pd}^2 S(S+1) N_{Mn}}{3k_B T_C} \langle s \rangle = \frac{\langle s \rangle}{\tilde{\chi}} \tag{9}$$

where $\tilde{\chi}$ denotes the susceptibility. Using equations (1) and (6) in (9) gives:

$$\varepsilon_\downarrow - \varepsilon_\uparrow = \frac{2\int (n_\uparrow(z) - n_\downarrow(z)) |\varphi(z)|^2 dz}{\tilde{\chi}} = \frac{2(N_\uparrow - N_\downarrow)\int |\varphi(z)|^2 |\psi(z)|^2 dz}{\tilde{\chi}\int |\psi(z)|^2 dz} = \frac{(N_\uparrow - N_\downarrow)}{\chi} \tag{10}$$

where $N_{\uparrow,\downarrow}$ denote 2D carrier densities as usual, $\psi(z)$ the non-equilibrium RTD wave function, the integrals are over the width of the well and $\chi$ is the 2D spin susceptibility computed in equations (6) and (8). For quantum wells in equilibrium $\psi(z) = \varphi(z)$; using equation (6) in (10), we trivially recover the familiar expression[8] for the Curie temperature in these systems: $T_C = \frac{J_{pd}^2 S(S+1) N_{Mn} \nu}{6k_B T_C} \int |\varphi(z)|^4 dz$. For large-barrier RTD's, the approximation $\psi(z) = \varphi(z)$ should still be valid in the non-equilibrium case. We therefore predict that the critical temperature should fall from this equilibrium quantum-well expression, to approximately half this value when the down-stream chemical potential falls below the quantum-well resonance energy.

## III. NUMERICAL SIMULATION

The simulated device consists of a highly p-doped GaAs emitter and collector, $Al_{0.8}Ga_{0.2}As$ barriers 10Å in thickness, and a 10Å thick (Ga,Mn)As well. We use the following material parameters, from Davies:[10] GaAs/AlGaAs *band-offset* – 0.38eV, *hole* *effective mass* – 0.52m$_0$ in GaAs and 0.55m$_0$ in AlGaAs, *relative permittivity* – 13.18 for GaAs and 10.68 for AlGaAs. We assume the GaAs parameter values for (Ga,Mn)As. As mentioned earlier, a quantitatively accurate calculation would have to take into account the more complicated valence band structure. The acceptor concentration in the emitter and collector is assumed to be 1x10$^{20}$cm$^{-3}$, while that in the barriers is assumed to be 2x10$^{19}$cm$^{-3}$. We assume a carrier (hole) concentration of 8x10$^{19}$cm$^{-3}$ in the well, after 80% compensation of the Mn due to anti-site defects in the GaAs. The emitter and collector are assumed to be sufficiently long to define the equilibrium chemical potential across the structure. The cross-section is assumed to be large enough for the transverse states to be treated as plane waves.

In the longitudinal (parallel to transport) direction $-z$, we employ a single-band nearest-neighbor tight-binding Hamiltonian $H_L$, written in a real space basis, following Datta.[6] Energies are measured with respect to the valence band edge in the emitter In this tight-binding model, the valence band offsets for the barriers are added to the energies at lattice points in the barriers. Therefore, we can write:

$$H_{L\downarrow,\uparrow}\left|z_n\right\rangle = \left(2t + \Delta E_V + U \pm h_{pd}/2\right)\cdot\left|z_n\right\rangle - t\cdot\sum_{\delta}\left|z_{n+\delta}\right\rangle \qquad (11)$$

$\Delta E_V$ is the valence band offset, $U$ is the electrostatic (Hartree) energy, $\Delta$ is a spin-dependent potential due to the magnetic impurities, $t \equiv \hbar^2/2m_l^* a^2$ is the hopping energy, $n$ is the site index and $\delta = \pm 1$. The mesh spacing $a$ used here is 1Å. The retarded Greens functions for spin-down and spin-up carriers are then given by:

$$G_{\uparrow,\downarrow}(E) = \left[E - H_{L\uparrow,\downarrow} - \Sigma_e - \Sigma_c\right]^{-1} \qquad (12)$$

where $E$ is the longitudinal carrier energy and $\Sigma_e$, $\Sigma_c$ are the self-energies corresponding to the emitter and collector. For the simple 1-D problem considered here, the self-energies are most easily evaluated by assuming outgoing plane wave boundary conditions.[6] The broadening functions for the emitter and collector are given by:

$$\Gamma_{e,c} = i\cdot\left(\Sigma_{e,c} - \Sigma_{e,c}^{\dagger}\right) \qquad (13)$$

The spectral function contributions from the emitter and collector are:

$$A_{e,c} = G\cdot\Gamma_{e,c}\cdot G^{\dagger} \qquad (14)$$

Obviously, the spectral functions and the Green functions above are implicitly spin-dependent. Because we neglect spin-orbit coupling, the equations for majority and minority spin spectral functions decouple. We assume constant chemical potentials $\mu_e$, $\mu_c$ deep inside the emitter and collector; $V = \mu_e - \mu_c$ is the applied bias. The non-equilibrium density matrices for spin-up and spin-down carriers are obtained from the NEGF expression:

$$\rho_{\uparrow,\downarrow} = \int_{-\infty}^{\infty} \frac{dE}{2\pi} \left[ F_0\left(E-\mu_e\right)\cdot A_{e\uparrow,\downarrow} + F_0\left(E-\mu_c\right)\cdot A_{c\uparrow,\downarrow} \right] \tag{15}$$

where,

$$F_0\left(E-\mu\right) = \sum_{\vec{\kappa}} f_0\left(E+\varepsilon_{\vec{\kappa}}-\mu\right) = S\cdot\frac{m_t^* k_B T}{2\pi\hbar^2}\cdot\ln\left[1+\exp\left(\frac{\mu-E}{k_B T}\right)\right] = S\cdot\nu\cdot\ln\left[1+\exp\left(\frac{\mu-E}{k_B T}\right)\right]$$

is the sum of the Fermi occupation probabilities for any one spin over all 2D transverse (perpendicular to transport) wavevectors $\vec{\kappa}$. Here, $S$ is the (large) transverse cross sectional area, and $\varepsilon_\kappa = \frac{\hbar^2\vec{\kappa}^2}{2m_t^*}$ are the plain wave energy eigenstates in the transverse direction. The carrier densities are then given by:

$$\Omega\cdot p_{\uparrow,\downarrow}\left(z\right) = \left[\rho_{\uparrow,\downarrow}\left(z,z'\right)\right]_{z'=z} \tag{16}$$

where $\Omega = a$, is in general, the volume of an individual cell. The Hartree potential energy $U$ appearing in the Hamiltonian is given by the Poisson equation:

$$\frac{d}{dz}\left(\varsigma\left(z\right)\frac{dU}{dz}\right) = q^2\cdot\left[p-N_A\right] = q^2\cdot\left[p_\uparrow + p_\downarrow - N_A\right] \tag{17}$$

$\varsigma$ being the dielectric function, $p_\uparrow$, $p_\downarrow$ the concentrations of spin-up and spin-down holes, and $N_A$ the acceptor ion concentration. The mean polarization of a magnetic ion in the absence of an external field within the mean-field picture is given by: [5]

$$\langle M\rangle_I = S\cdot B_S\left(J_{pd}\cdot S\cdot\left[p_\uparrow(R_I) - p_\downarrow(R_I)\right]/2k_B T\right) \tag{18}$$

where $B_S$ is the Brillouin function:

$$B_S(x) = \frac{2S+1}{2S}\coth\left(\frac{2S+1}{2S}\cdot x\right) - \frac{1}{2S}\coth\left(\frac{1}{2S}\cdot x\right) \approx \frac{S+1}{3S}\cdot x, x \ll 1 \tag{19}$$

Here $J_{pd}$ is assigned a value of 150meV.nm$^3$, in the same range as found experimentally.[8]

Then the spin-dependent kinetic-exchange potential is obtained, in the continuum limit, from:

$$h_{pd}(z) = J_{pd} \cdot N_{Mn}(z) \cdot \langle M \rangle (z) \tag{20}$$

The set of equations (11) through (16), describe transport and magnetotransport properties of this system when solved self-consistently with (17) and (20). The boundary condition for the Poisson equation that is the best suited for nanostructure simulation is Neumann.[11] The Neumann problem for the Poisson equation is singular but that does not pose a hindrance here, as we solve it self-consistently with the quantum kinetic equations by a Newton-Raphson iteration scheme.[12] The iterative solution for the spin-dependent potential energy proceeds by a bisection scheme.

The voltage-dependent critical temperature can be calculated following the spirit of the analytical approach of Lee et al.[8] A small spin-splitting is introduced in the paramagnetic state close to $T_C$, and the resulting carrier density difference is calculated using equation (20) and the linear expansion in equation (19) to obtain:

$$p_{\uparrow}(z;V) - p_{\downarrow}(z;V) = \frac{6k_B T_C}{S(S+1)J_{pd}} \cdot \langle M \rangle (z;V) = \frac{6k_B T_C}{S(S+1)J_{pd}^2 N_{Mn}} \cdot h_{pd}(z;V)$$

$$\Rightarrow L(\Delta(z;V)) = p_{\uparrow}(z;V) - p_{\downarrow}(z;V) = \frac{6k_B T_C}{S(S+1)J_{pd}^2 N_{Mn}} \cdot h_{pd}(z;V) \tag{21}$$

$$\Rightarrow T_C(V) \approx \frac{S(S+1)J_{pd}^2 N_{Mn}}{6k_B} \left( \frac{\delta L(V)}{\delta \Delta} \right)_{z=\bar{z}, \Delta \to 0} \tag{22}$$

where $\bar{z}$ denotes the center of the quantum well, and, $L$ maps a small $\Delta(z)$ to $p_{\uparrow}(z) - p_{\downarrow}(z)$ through a solution of the quantum transport and Poisson equations. We assume,[8] that $\Delta(z) \propto |\psi(z;V)|^2 \approx c.p(z;V)$, $c \to 0$ and consider the linear variation of $p_{\uparrow}(\bar{z}) - p_{\downarrow}(\bar{z})$ as a function of $\Delta(\bar{z})$ (by letting $c$ increase in steps) for each $V$. The critical temperature is proportional to the slope of this curve as seen from equation (22).

Figure 2 plots results for the critical temperature $T_C$ versus applied voltage evaluated in this way for two sets of barrier heights and widths. The spline-fitted solid curve corresponds to the GaAs/AlGaAs system specified earlier, with a barrier height $V_0 = 0.38eV$ and width $b = 1.0nm$. For the parameters used here, the simple theory of Section II predicts steps in the $T_C$ – V characteristics at $v_S \cong 0.07V$ and $v_C \cong 0.23V$. The steps in $T_C$ are not abrupt but smear out over a range of applied bias. This spread is primarily due to the finite width of the resonance. To illustrate the resonance-width effect, we have performed a second simulation with a taller and wider barrier, and therefore a narrower resonance. We see that the width of the steps in the $T_C$ – V curve is indeed reduced as the resonance becomes sharper. Figures 3(a), (b), (c) show the valence band edge, the carrier concentration profile and the transmission probability, in the paramagnetic state at an applied bias of 25mV; Figures 4(a), (b), (c) show similar plots for 125mV. They illustrate the two regimes $\mu_L > E > \mu_R$ and $\mu_L > \mu_R > E$ – a transition from the one to the other leads to the first step in the $T_C$ – V plot. Note that the carrier concentration in the well barely changes in going from one region of operation to the other.

We performed a 20K simulation, where a very high degree of spin polarization is to be expected at equilibrium – the experimental $T_C$ for bulk (Ga,Mn)As being about 110K.[2] Here we self-consistently solved the equations (15), (17) and (20) for the coupled transport, electrostatic, and magnetic states as explained earlier. The I – V characteristics for the RTD, in figure 5, show strongly spin-polarized currents at low bias, and also indicate that the magnetism vanishes around 300mV. From the analytic theory, we expect vanishing magnetization to occur when the resonance is pulled down below the band-edge on the emitter side. At this point, the current should fall sharply giving rise to the well-known negative differential resistance (NDR) regime of the RTD; this is indeed seen to be the case in Fig.5. A plot of the spin polarization $\alpha = \left(p_\uparrow - p_\downarrow\right)/\left(p_\uparrow + p_\downarrow\right)$ along the device shown in figure 7, clearly illustrates the decay of the magnetization with applied bias.

Simulations identical to those described in the previous paragraph were performed for varying temperatures to obtain the full 3D plot, seen in figure 7, of the peak spin polarization $\alpha_{\max}$ inside the well as a function of both temperature and voltage. The contour $\alpha_{\max} = 0 \Rightarrow T = T_C\left(V\right)$ is in agreement with the $T_C$ – V characteristic plotted in figure 2.

## IV. CONCLUSIONS

In this paper we have considered the coupling between magnetic and transport properties of a resonant tunneling diode (RTD) with a diluted-magnetic-semiconductor quantum well. The large literature on RTD structures has generally focused on the region of negative differential conductance which occurs when the quantum well resonance energy falls below the *emitter* band-edge. In this paper we predict that the ferromagnetic transition temperature of the quantum well system decreases abruptly by approximately a factor of two when the Fermi level of the *collector* falls below the quantum well resonance energy, a condition that in typical quantum well structures holds over a broad interval of bias voltages below the negative differential conductance regime. The ferromagnetic transition temperature of the quantum well is expected to drop to near zero in the negative differential conductance regime.

The voltage dependence of the critical temperature and of the spin-polarization at a given temperature that is predicted here could be verified by simply measuring the RTD I-V characteristics or more directly by magnetic or optical circular dichroism[13] measurements. The prediction of a bias voltage dependent magnetic phase transition temperature is potentially interesting for applications. For example the three step $T_C - V$ characteristic might possibly suggest potential as a three-level logic device. More practically perhaps, this novel effect could also, when operated at an appropriate temperature, allow a small bias voltage to change the quantum well system between magnetic and non-magnetic states.

There are a number of practical difficulties that would have to be overcome for this idea to be tested experimentally, as well as a number of theoretical issues that require further attention. On the materials and experiment side, the geometry we have considered, in which the magnetic layer is the quantum well of a RTD, requires that AlGaAs be grown on top of (Ga,Mn)As *at low-temperatures*, in order to avoid precipitation of Mn and MnAs intermetallic compounds. Low temperature growth is known to lead to low quality interfaces[14] and will no doubt lead to inhomogeneous broadening of the sharp quantum well resonance that is responsible for the relatively abrupt decrease in the ferromagnetic transition temperature in our models. Verification and refinement of the effect we propose would have to go hand in hand with advances in epitaxial growth techniques in (Ga,Mn)As/AlGaAs systems in order to realize the abrupt transition temperature changes that occur in our simple model. On the theoretical side, our predictions would be altered somewhat by using a more realistic model for the semiconductor valence bands. This kind of improvement in the theory is readily implemented. More challenging is the analysis of the importance of corrections to the mean-field theory that we employ. For a single parabolic band system, a two-dimensional (Ga,Mn)As ferromagnet has vanishing spin-stiffness, implying that corrections to mean-field theory have an overriding importance.[15] For a valence band system, however, the finite well width creates a large anisotropy gap and corrections to the mean-field theory are likely to be less important.

## ACKNOWLEDGEMENTS

This work was supported in part by the MARCO MSD Focus Center NSF, and the Texas Advanced Technology Program. AHM was supported by the Welch Foundation, by the Department of Energy under grant DE-FG03-02ER45958, and by the DARPA SpinS program. Swaroop Ganguly would like to thank Prof. Supriyo Datta for helpful discussions.

REFERENCE

[*] Electronic Address: swaroop@ece.utexas.edu

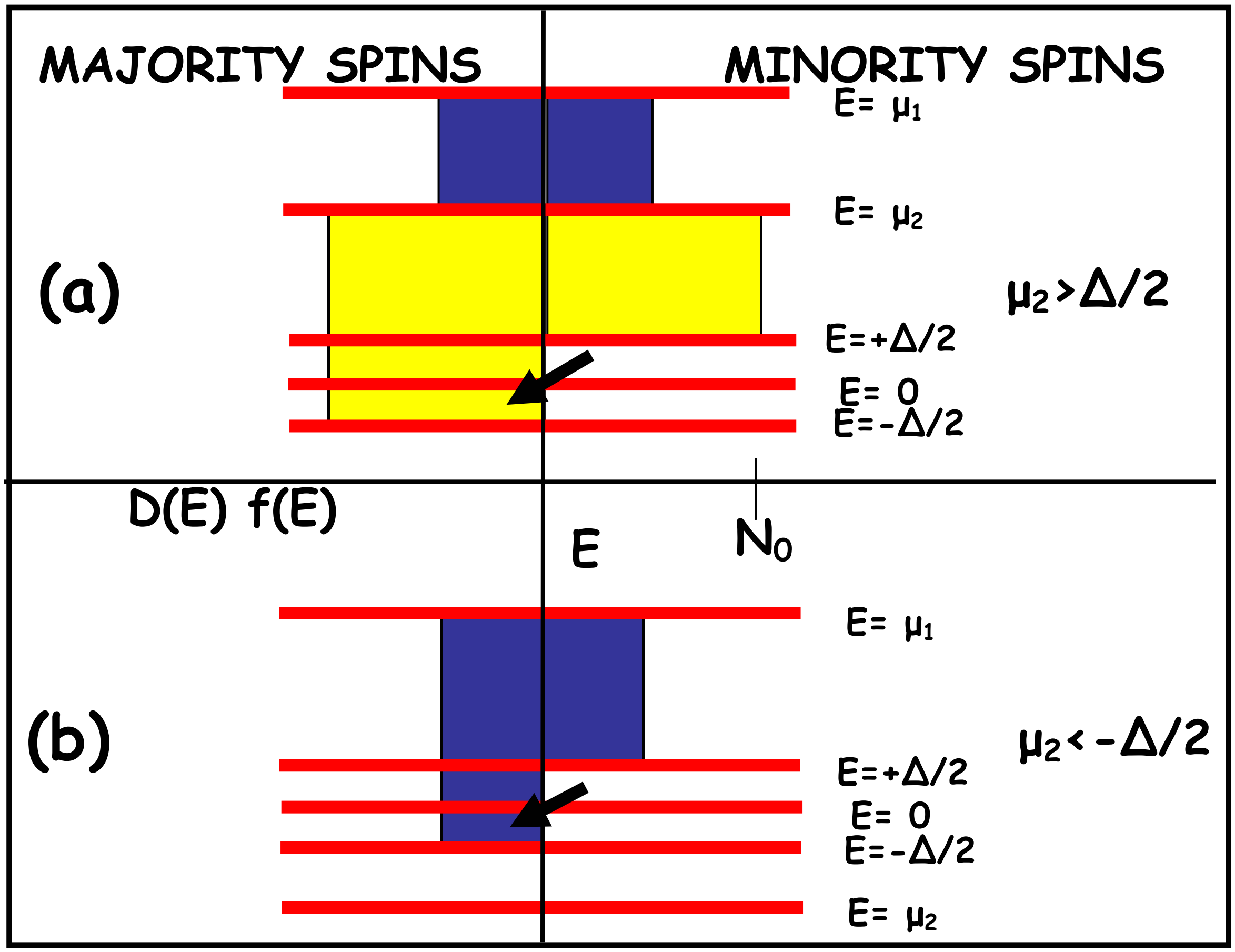

Fig. 1 (Color online) Non-equilibrium majority and minority spin occupation numbers for spin-polarized bands. In mean-field theory the critical temperature is proportional to the magnetic susceptibility, which is proportional to the reduction in band energy due to a small Zeeman spin-splitting energy Δ. In (a) the quantum-well resonance energy lies below the down-stream chemical potential. The band energy gain due the spin-splitting has the same value as in equilibrium. In (b) the quantum well resonance energy lies above the down-stream chemical potential. The band energy gain due to Zeeman spin-splitting is then half as large as in the equilibrium case.

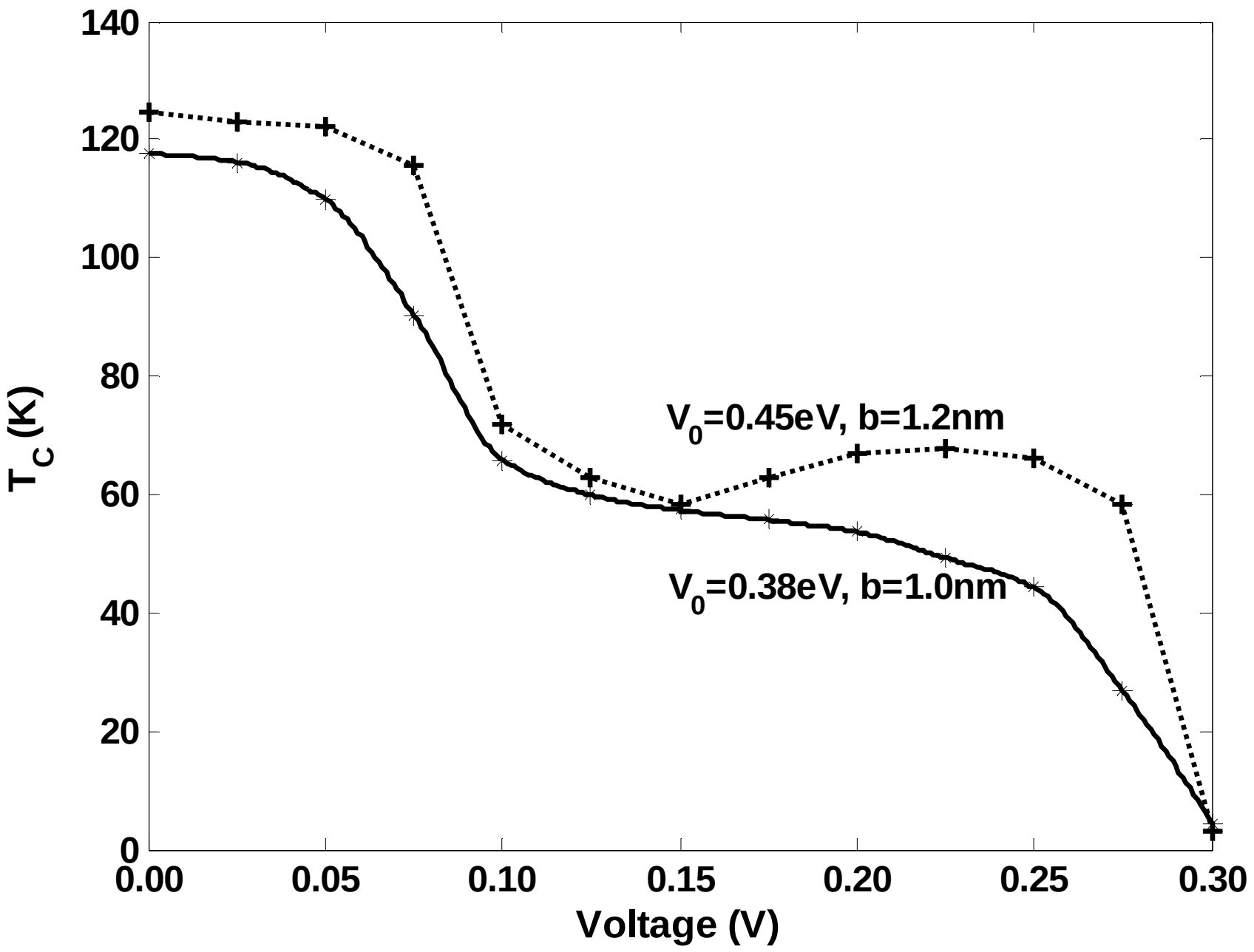

FIG.2. Numerically simulated $T_C$ – V characteristics for different barrier heights and widths. Notice that the critical temperature variation is more abrupt when the barrier is higher and thicker, and the quantum well resonance is narrower.

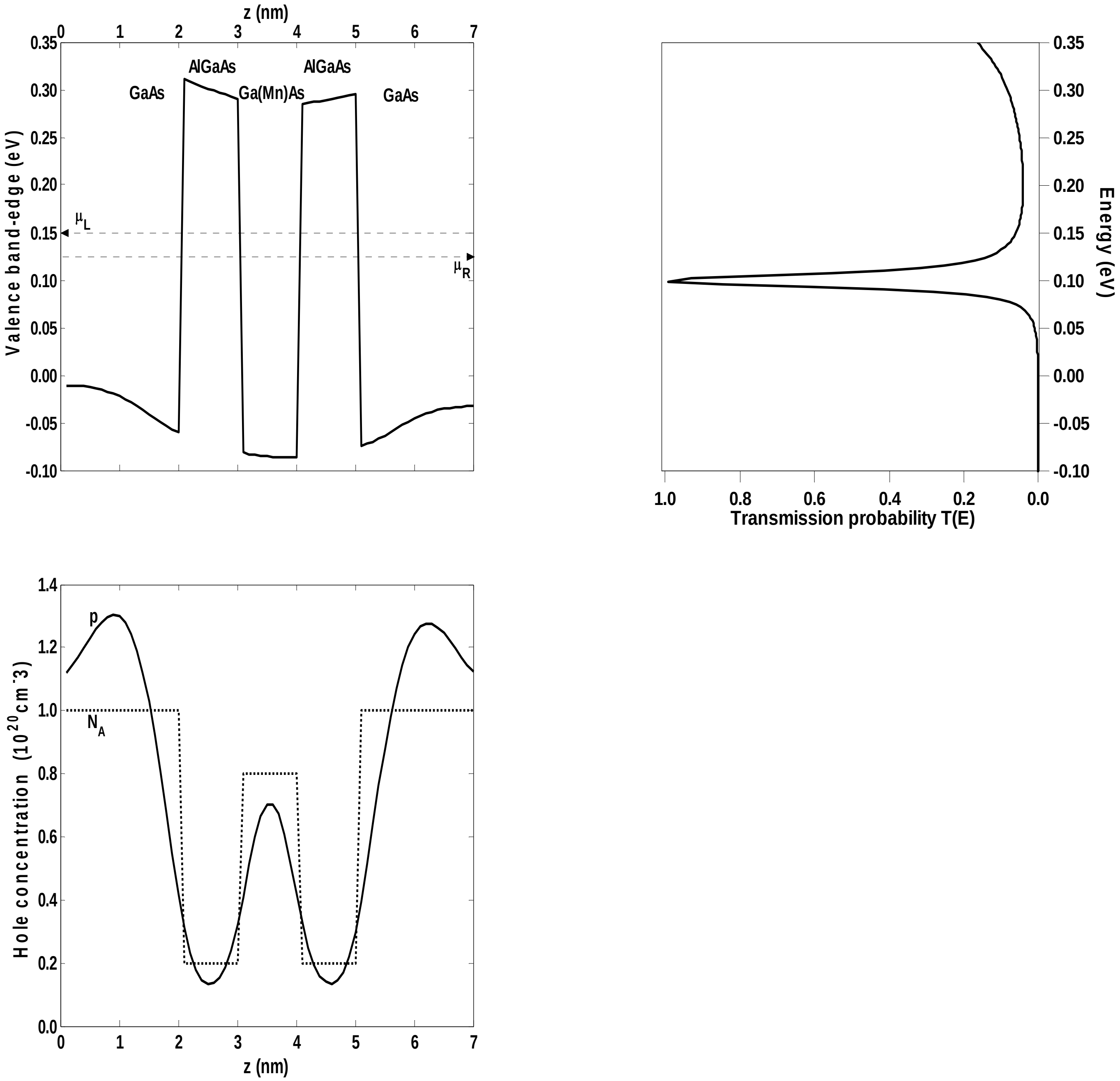

FIG. 3(a), (b), (c). Paramagnetic state: V=25mV; $\mu_L$ , $\mu_R$ are the chemical potentials in the left and right leads; p is the hole concentration and $N_A$ the acceptor doping profile.

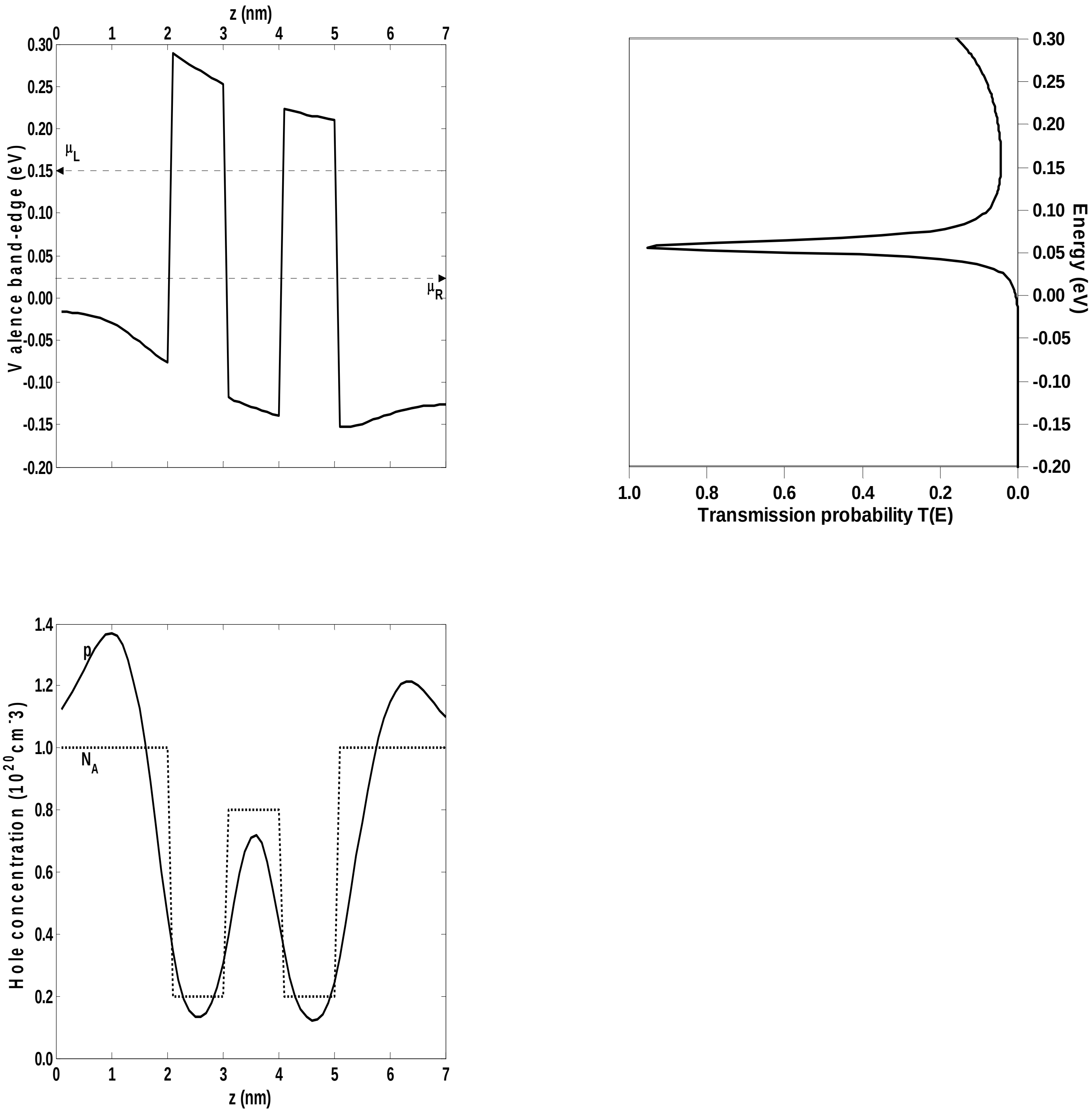

FIG. 4(a), (b), (c). Paramagnetic state: V=125mV; $\mu_L$ , $\mu_R$ are the chemical potentials in the left and right leads; p is the hole concentration and $N_A$ the acceptor doping profile.

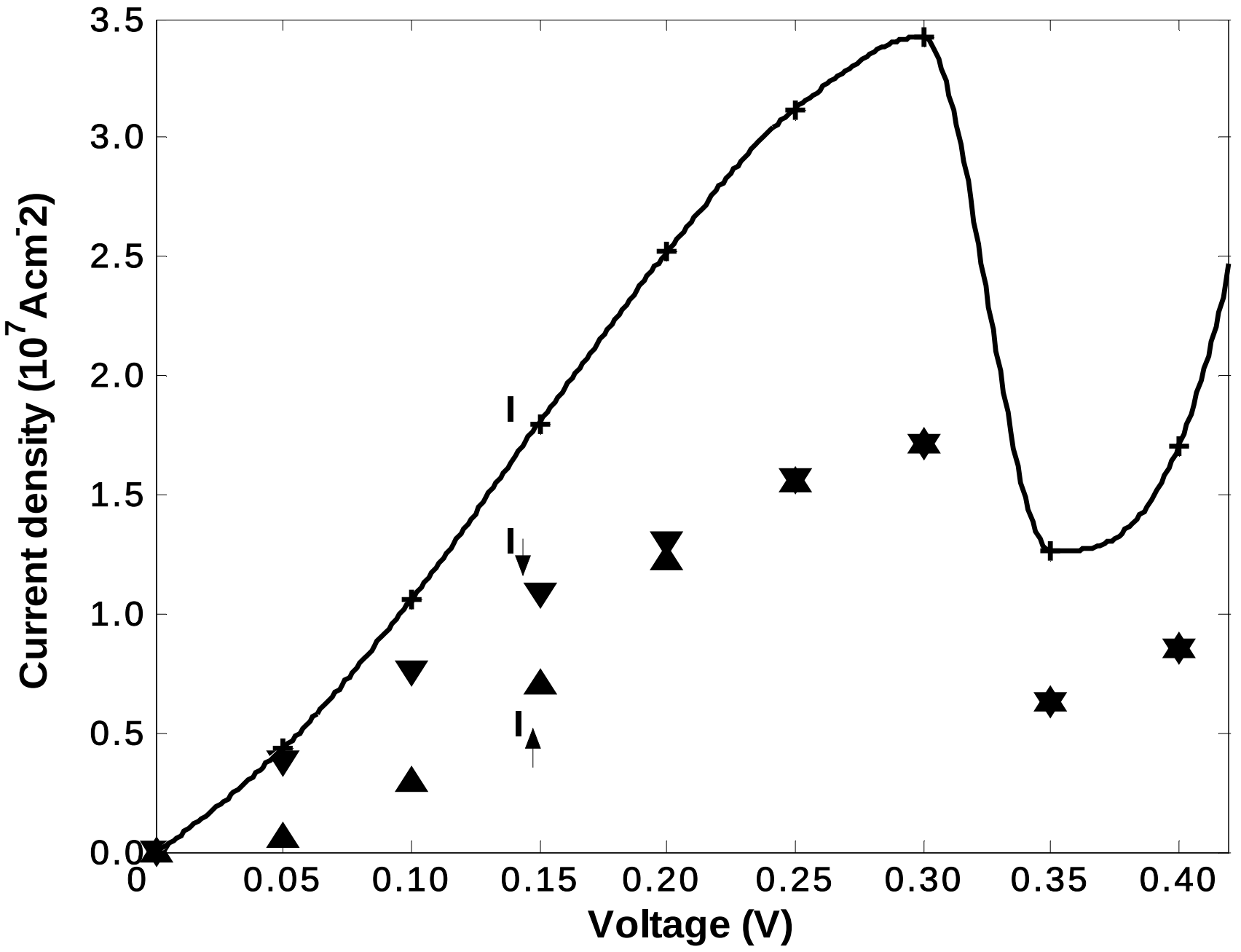

FIG. 5. I-V characteristics of RTD (T=20K) showing total current (solid line), majority spin current (triangle-up) and minority spin current (triangle-down).

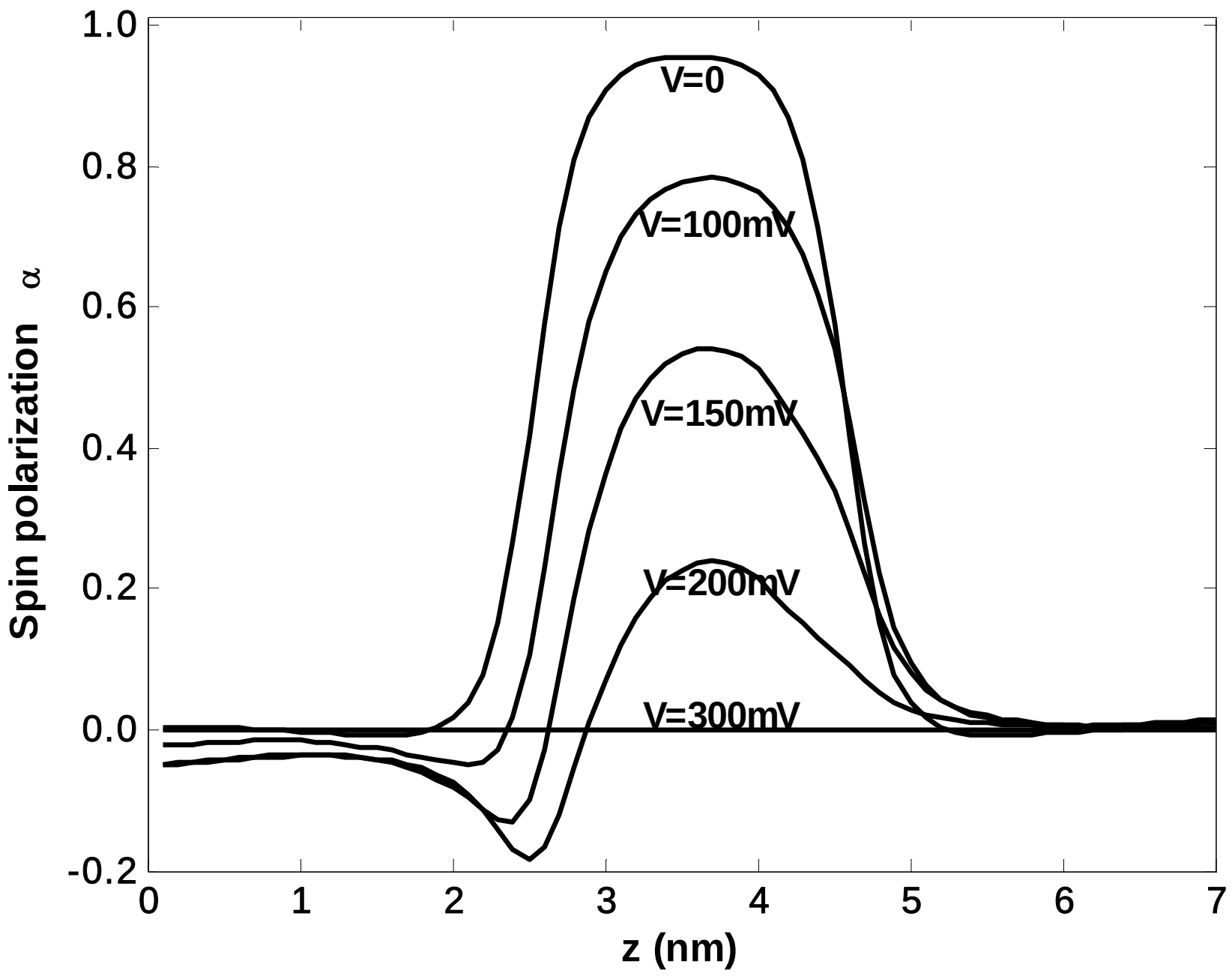

FIG. 6. Spin polarization along the length of the device (T=20K) for increasing applied voltage.

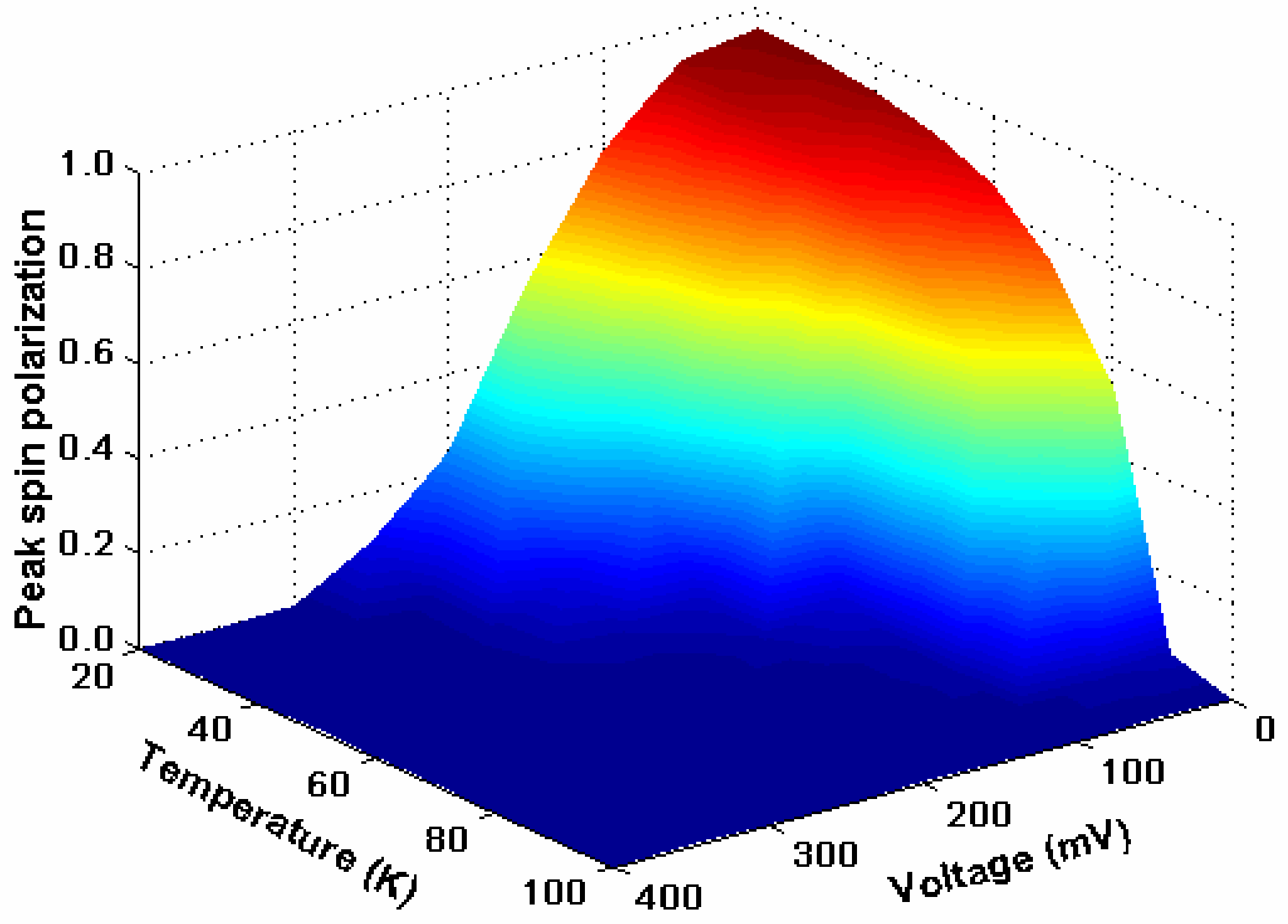

FIG. 7. (Color online) Peak spin polarization vs. voltage and temperature.